\documentclass[aps,prb,twocolumn]{revtex4-1}

\usepackage{graphicx}
\usepackage{amsmath}
\usepackage{dcolumn}

\usepackage{color}

\newcommand{\Eq}[1]{Eq.~\eqref{#1}}
\newcommand{\Eqs}[1]{Eqs.~\eqref{#1}}

\begin{document}
\title{Period tripling subharmonic oscillations in a driven superconducting resonator}

\author{Ida-Maria \surname{Svensson}}
\email[]{ida-maria.svensson@chalmers.se}
\author{Andreas \surname{Bengtsson}}
\author{Philip \surname{Krantz}}
\author{Jonas \surname{Bylander}}
\author{Vitaly \surname{Shumeiko}}
\author{Per \surname{Delsing}}
\email[]{per.delsing@chalmers.se}
\affiliation{Microtechnology and Nanoscience, Chalmers University of Technology, SE-41296 G\"{o}teborg, Sweden}
\date{\today}

\begin{abstract}
We have observed period-tripling subharmonic oscillations in a driven superconducting coplanar waveguide resonator operated in the quantum regime, $k_B T \ll \hbar\omega$. The resonator is terminated by a tunable inductance that provides a Kerr-type nonlinearity. We detected the output field quadratures at frequencies near the fundamental mode, $\omega/2\pi \sim 5\,$GHz, when driving the resonator with a current at $3\omega$, with amplitude exceeding an instability threshold. We observed three stable radiative states with equal amplitudes, phase-shifted by $2\pi/3$ radians, red-detuned from the fundamental mode. The downconversion from $3\omega$ to $\omega$ is strongly enhanced by near-resonant excitation of the second mode of the resonator, and the cross-Kerr effect. Our experimental results are in quantitative agreement with a model for the driven dynamics of two coupled modes.
\end{abstract}

\maketitle

\section{Introduction}

Nonlinear dynamical systems exhibit a vast variety of behaviors, from simple effects such as harmonic generation to sophisticated multiple bifurcations, to pattern formation, and chaos \cite{Strogatz,Cross1993,Sagdeev}. Particularly interesting are strongly nonlinear phenomena in the quantum regime, which can be realized in low-dissipative microwave systems such as circuit quantum electrodynamics (cQED) devices. Such phenomena play a central role and are widely employed in quantum information technology for qubit read-out \cite{Siddiqi2004,Lin2014,Krantz2016},  photon entanglement \cite{Eichler2011,Bergeal2012,Flurin2012,Macklin2015} and generation of Schr\"odinger cat states \cite{Mirrahimi2014,Puri2017}.

Period-multiplying subharmonic oscillations \cite{Hayashi,JordanSmith} constitute a particular class of nonlinear phenomena. The oscillations appear as a nonlinear response at the oscillator frequency to an external drive at a multiple of the resonant frequency.  In the quantum picture, the elementary process that underlies the subharmonic oscillations is a decay of a single photon into three, four, or more photons. The subharmonic oscillations are described by non-perturbative solutions to the dynamical equations, which appear abruptly and co-exist with the stable vacuum state. In this respect the subharmonic oscillations  distinctly differ from conventional parametric oscillations, which gradually emerge as a result of the vacuum instability. This difference is analogous to the difference between a first and a second order phase transition \cite{Haken1975}. Furthermore, a symmetry breaking aspect of this difference has important implications for the quantum dynamics of the period-tripling oscillations \cite{Zhang2017}.

Although the period-multiplying phenomenon is theoretically explained in textbooks, experimental demonstrations are not common. A few early observations of subharmonic resonances in electromagnetic devices were performed on essentially classical electrical circuits with saturable inductors \cite{Spitzer1945} or varactors \cite{Linsay1981}. More recent reports concern subharmonic resonances in lasers \cite{Ngai1993,SotoCrespo2004}. In Josephson circuits, the period-multiplying phenomenon has not received much attention; instead, research was focused on transition to chaos  \cite{Huberman1980,Kautz1981} and lately on bifurcation phenomena \cite{Siddiqi2005,Vijay2009} and parametric oscillations \cite{Wilson2010,Wustmann2013,Lin2014,Krantz2016,Yamamoto2016}. Only recently, subharmonic oscillations in the quantum regime were theoretically discussed in the context of cQED \cite{Guo2013,Denisenko2016,Zhang2017}.

In this paper we report the first experimental observation of period-tripling subharmonic oscillations in a driven superconducting resonator in the quantum regime, $k_BT\ll \hbar\omega$ \cite{Sandberg2008,Wallquist2006,PalaciosLaloy2008}, where the thermal energy is much smaller than the energy of a single photon. We drive the nonlinear resonator with a harmonic signal, with power $P_d$, at frequency $3\omega$, approximately equal to three times the fundamental resonator mode frequency, and observe a strong response at $\omega$. The output microwave signal consists of three correlated beams with equal amplitudes and different phases, shifted by $2\pi/3$ radians. The oscillations are detected within a certain window of the driving field amplitude: they start at finite threshold detuning within the resonator bandwidth and persist deep into the red detuning region. 

Our observations can be qualitatively understood from the theory of a nonlinear oscillator \cite{Hayashi,JordanSmith}. When a driving force with off-resonant frequency $3\omega$ is applied, it generates a linear response at the same frequency, which is downconverted to frequency $\omega$ due to nonlinearity. The downconversion has the highest efficiency when the  detuning, $\delta_1= \omega-\omega_1$, from the oscillator resonant frequency $\omega_1$ is small, $\delta_1 \ll \omega_1$. 

However, application of this single-mode scenario to the resonator setting requires additional considerations. 
In our experiment, we drive the resonator close to its second mode, $\omega_2$, such that $\delta_2 = 3\omega - \omega_2 \ll \omega_1$. As a result, the response at the driving frequency becomes strongly enhanced and nonlinear, and the system dynamics is well described by two strongly interacting modes. This situation is different from the single mode oscillator model explored in previous works on subharmonic oscillations. In fact, this resonant enhancement of the external drive, by more than three orders of magnitude, is crucial for the possibility to access the subharmonic oscillation regime in experiments.

\section{Experiment}

\begin{figure}
	\center
	\includegraphics{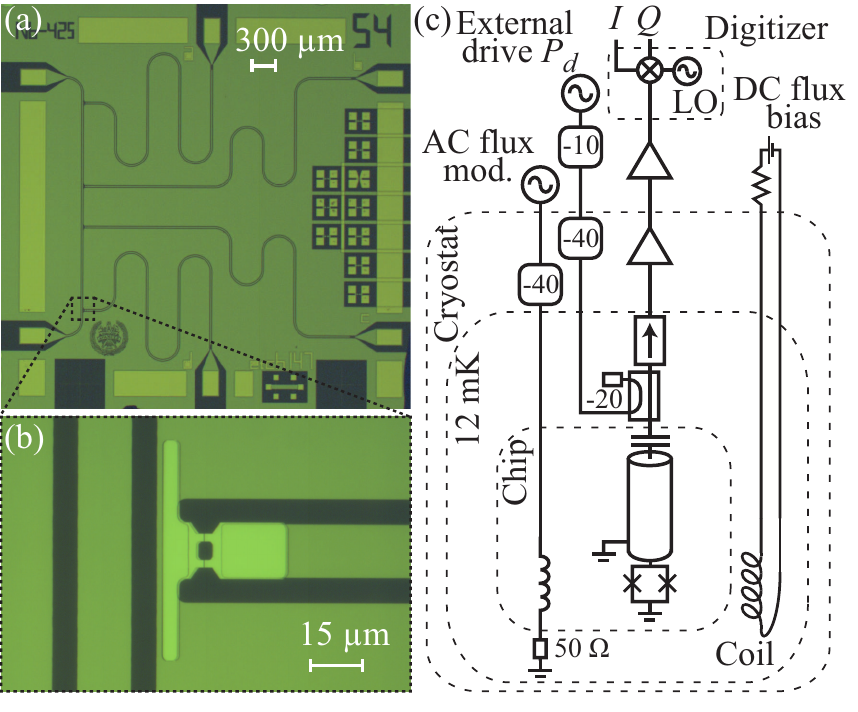}
	\caption{\label{fig:SampleAndSetup}(a) Optical micrograph of one of the samples. The four coplanar waveguide resonators meander between the on-chip flux line on the left and the contact pads on the right. (b) Zoom-in on the SQUID that terminates the bottom resonator. The SQUID is designed with two identical Josephson junctions. (c) Measurement setup. Microwave signals are applied via attenuated coaxial cables, one for direct driving by an external AC current, and one for flux modulation. To separate input and output, a directional coupler is used to route the signal. A static magnetic flux is induced by a superconducting coil. The output signal is amplified in a $4-8\,$GHz bandwidth by a cryogenic amplifier as well as a room temperature amplifier. The quadrature voltages are acquired by heterodyne detection followed by digital demodulation.}
\end{figure}
To observe subharmonic oscillations we use a frequency-tunable coplanar waveguide microwave resonators\cite{Sandberg2008,Wallquist2006,PalaciosLaloy2008}. The resonator is capacitively coupled to a $50\,\Omega$ transmission line in one end, and grounded via a superconducting quantum interference device (SQUID) in the other end, see Fig.~\ref{fig:SampleAndSetup}. In practice, the SQUID acts as a tunable nonlinear inductance controlled by the magnetic flux $\Phi$ threading its loop and the current $I_s$, $L_J(\Phi_{},I_s)=\Phi_0/\left(2\pi|\cos(\pi\Phi_{}/\Phi_0)|\sqrt{I_c^2-I_s^2}\right)$, where $\Phi_0=h/2e$ is the magnetic flux quantum, and $I_s$ and $I_c$ denote the current flowing through the device and its critical current, respectively. The Josephson nonlinearity of the SQUID induces, at weak excitation, a Duffing-Kerr nonlinearity in the resonator.

The samples are fabricated using standard processes: Josephson junctions are deposited by two-angle evaporation of aluminum. The rest of the circuit is etched out of a sputtered layer of niobium on a sapphire wafer. A micrograph of one chip with four resonators of different lengths is shown in Fig.~\ref{fig:SampleAndSetup}(a). The chip is anchored to the mixing chamber stage of a dilution refrigerator with a base temperature of $12\,$mK.

The measurement setup is sketched in Fig.~\ref{fig:SampleAndSetup}(c). For static magnetic flux biasing of the SQUID we use a superconducting coil mounted close to the sample box. The SQUID nonlinearity can be modulated by applying a microwave signal as an external drive of the current through the SQUID $I_s$ or by flux modulation. We focus mainly on the external driving.  The resonator output signal is amplified by a low-noise cryogenic amplifier at the $3\,$K stage, as well as an additional amplifier at room temperature, before being sampled by a digitizer. To maintain phase coherence, a $10\,$MHz signal is used to lock the signal generator and the digitizer together. The digitizer downconverts both the in- and out-of-phase quadratures, $I(t)$ and $Q(t)$, with a local oscillator before digitizing the data at an effective sampling rate $f_s$ during a time $t_s$. From the individual quadratures the total output power after amplification can be calculated as $P_{out}=\langle I^2\rangle+\langle Q^2\rangle$. 

\begin{table*}
	\caption{\label{tab:resonators} Resonator and SQUID parameters: $d$, resonator length; $I_c$, SQUID critical current; $C_J$, SQUID capacitance; $\gamma_0=L_J^0/L_0d$, the inductive participation ratio; $L_0$ and $C_0$, inductance and capacitance per unit length; $\omega_1(0)$, resonant frequency of the first mode at $\Phi=0$; $3\omega_1(0)-\omega_2(0)$, spectrum anharmonicity; $2\Gamma_1(0)$, damping of the first resonator mode; $Q_{1,int}$ and $Q_{1,ext}$, quality factors of the fundamental mode.}
\begin{center}
\begin{tabular}{|c|c|c|c|c|c|c|c|c|c|c|}
	\hline
	$d$ & $I_c$ & $C_J$ & $\gamma_0$ & $L_0$ & $C_0$ & $\omega_1(0)/2\pi$ & $(3\omega_1(0)-\omega_2(0))/2\pi$ & $2\Gamma_1(0)/2\pi$ & $Q_{1,int}(0)$ & $Q_{1,ext}(0)$\\ 
	$[\mu$m] & [$\mu$A] & [fF] & $[\%]$ & [$\mu$H/m] & [nF/m] & [GHz] & [MHz] &[MHz] & [$10^3$] & [$10^3$]\\ \hline
	5080 & $1.90$ & $86.1$ & $7.7$ & $0.44$ & $0.16$ & $5.504$ & 136 & 0.38 & 61.1 & 19 \\ \hline
\end{tabular}
\end{center}
\end{table*}

\begin{figure}
	\center
	\includegraphics{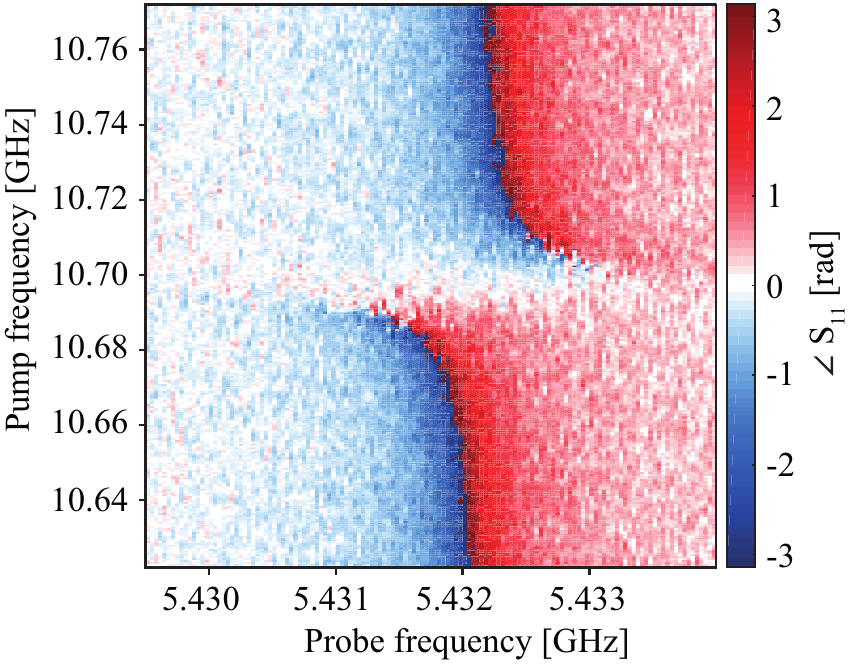}
	\caption{\label{fig:BSexample}
		Determination of the second mode frequency via parametric up-conversion. A weak probe tone (horizontal axis), near the resonant frequency of the first mode $\omega_1$, is up-converted by the flux pump tone (vertical axis), near the difference frequency $\omega_2-\omega_1$. The process results in an avoided level crossing visible in the reflected probe tone. The data presented shows the reflected phase response for a magnetic flux bias, $\Phi=0.17\,\Phi_0$, where $\omega_1/2\pi$  = 5.432$\,$GHz and $(\omega_2 - \omega_1)/2\pi$ = 10.70$\,$GHz, yielding $\omega_2/2\pi$  = 16.132$\,$GHz.
	}
\end{figure}

We directly probe the first resonator mode in a reflection measurement. Higher modes can only be measured indirectly due to the 4-8$\,$GHz bandwidth limitation of our setup. To detect the second mode we use parametric up-conversion \cite{ZakkaBajjani2011,Wustmann2017}: we modulate the magnetic flux penetrating the SQUID loop at the difference frequency of the first and the second modes, $\omega_2-\omega_1$, while simultaneously applying a weak drive tone at $\omega_1$. The flux pump converts photons from the first to the second mode, resulting in an avoided crossing, as shown in Fig.~\ref{fig:BSexample}, from which we can determine the difference frequency. 

\begin{figure}
	\center
	\includegraphics{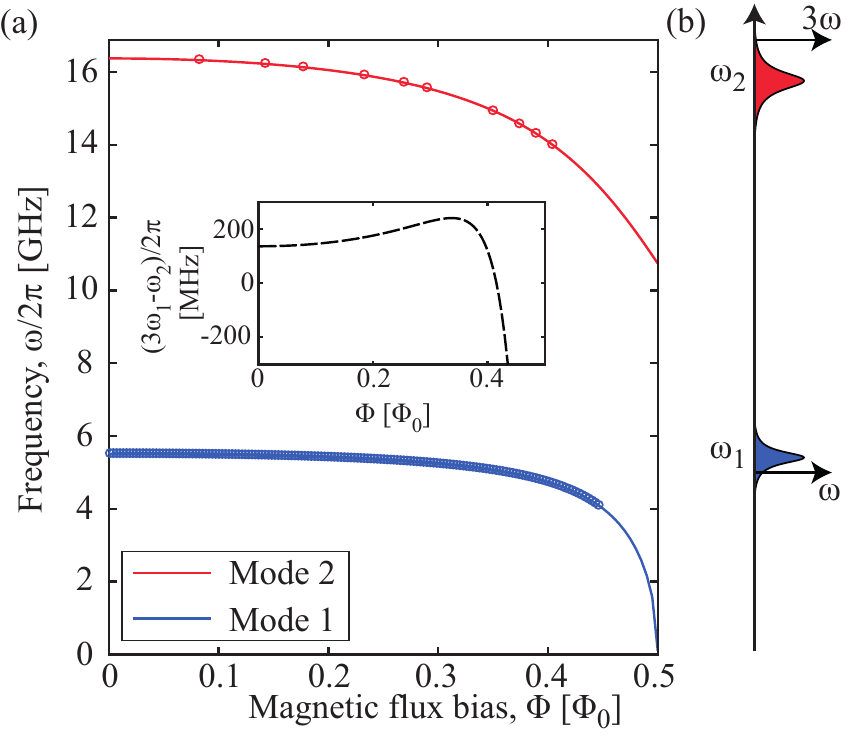}
	\caption{\label{fig:DCtuning}(a) Spectroscopy of resonator modes 1 and 2, fitted using \Eq{spectrum}.  Extracted resonator and SQUID parameters are found in Table \ref{tab:resonators}. Inset: anharmonicity of the spectrum, $3\omega_1-\omega_2$. (b) An illustration of the frequency spectrum. Since the spectrum is non-equidistant the drive signal around $3\omega$ falls slightly above mode 2 while the measurement frequency $\omega$ is slightly below mode 1.}
\end{figure}

In Fig.~\ref{fig:DCtuning}(a) we present the extracted frequencies of the lowest mode for our sample together with a fit to the spectral dispersion, \Eq{spectrum}. The extracted parameters are presented in Table~\ref{tab:resonators}. The resonator is overcoupled and have a narrow bandwidth, less than $1\,$MHz.

\section{Subharmonic oscillations}
\subsection{Observations}

\begin{figure*}
	\center
	\includegraphics{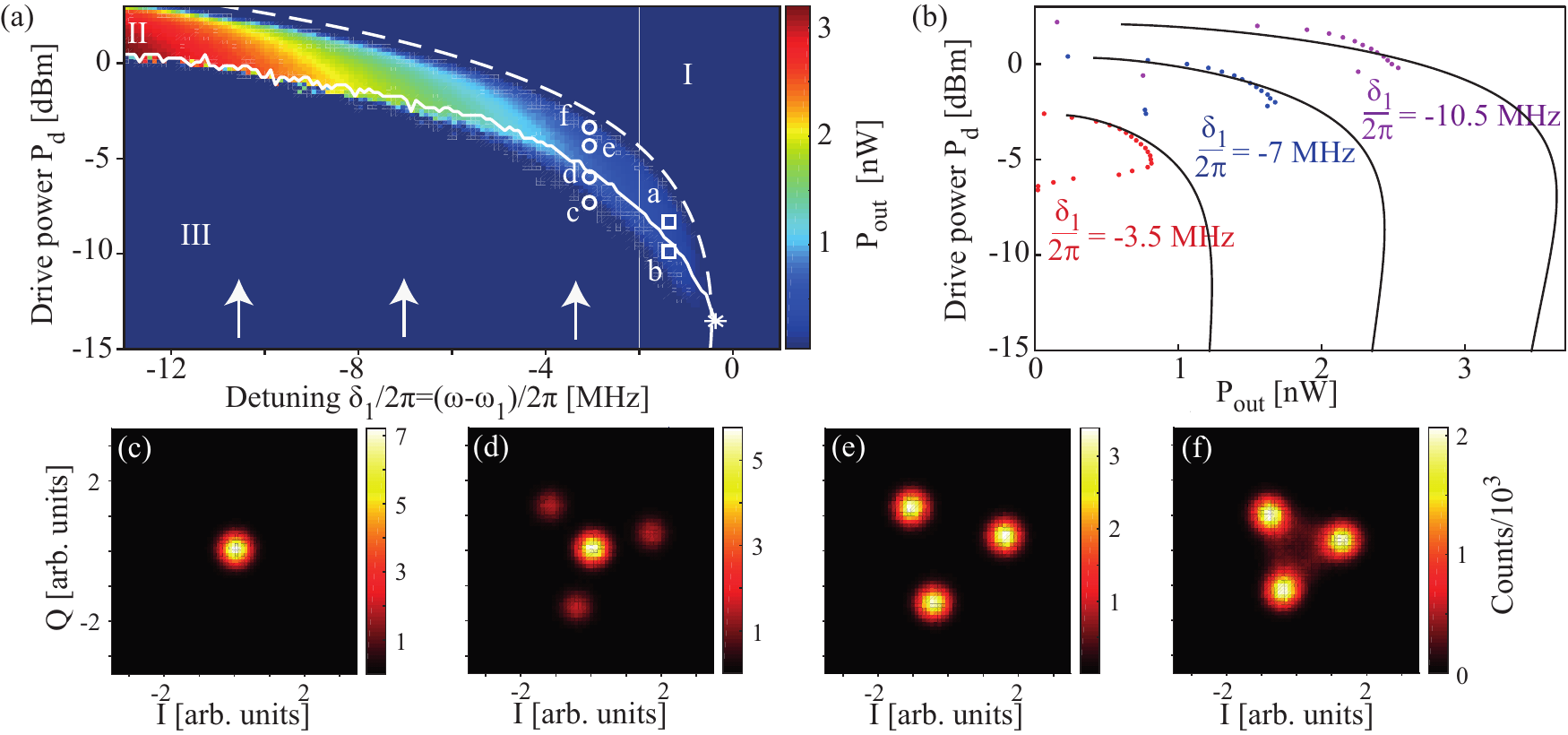}
	\caption{\label{fig:Region}
		(a) Intensity of the subharmonic oscillation output signal as function of drive power and detuning, for $\Phi_{}=0$. The oscillations are detected in region II. The white line corresponds to the maximum output signal for each measurement frequency; the white star at the end of the white line represents the subharmonic oscillation frequency threshold. The dashed white line that separates regions I and II corresponds to the theoretical boundary of existence for the subharmonic oscillations. In region III, the oscillations, although existing as a solution to Eqs.~\eqref{EOMtext}, are not visible because the oscillator switches to the ground state. (b) The dots represent three linecuts from (a), at detuning values indicated by the white arrows; a theory fit to  these three data traces is represented by the black lines. (c-f) Histograms of the detected radiation quadrature voltages from the operating points indicated by the white circles in (a). These histograms reveal three dynamic states: (c) the ground state, \textit{i.e.} the background noise level; (d) the ground state (in the middle) and the three excited states with equal amplitudes and with phases differing by $2\pi/3$; (e) and (f) the three excited states. All histograms are sampled at a rate of $f_s=100\,$kHz.  The features in (f) are connected by faint lines, an artifact from the averaging that indicates that the switching rate at this point is higher than in (d) and (e), and not negligible compared to $f_s$. These histograms can be compared to the histograms sampled with $f_s=10\,$kHz that are presented in Fig.~\ref{fig:fs10kHz}, and were measured with parameters as indicated by the white squares in (a).}
\end{figure*}

We observe period-tripling subharmonic oscillations at $\omega \approx \omega_1$, by applying an external signal at $3\omega$. This effect has been observed in several samples though in the paper we only present data for one sample. As can be seen in Fig.~\ref{fig:Region}(a), we observe subharmonic oscillations in a range of red detuning, $\delta_1=\omega-\omega_1<0$, and for signal generator drive powers above $\sim-13\,$dBm. This range of drive powers and detunings forms region II, where the subharmonic oscillations are visible. Above and below, in regions I and III, respectively, no subharmonic oscillations are observed.

We also investigated the quadratures of the oscillations using histograms of the $I(t)$ and $ Q(t)$ signals. In Fig.~\ref{fig:Region}(c) we show a background histogram (peaked at $ I= Q =0$) illustrating the system noise level. This histogram represents the system ground state. At higher drive power, inside region II, the histograms feature three well-defined stable states forming a regular triangle, see Fig.~\ref{fig:Region}(e) and (f). At the low-power edge of region II the system shows four states, see Fig.~\ref{fig:Region}(d). This observation is in full agreement with the phase portrait of the subharmonic oscillator, Fig.~\ref{phaseportrait}, featuring four coexisting stable steady states: the silent ground state and the three excited states. 

The system switching rates between the states are different for different operating points, $\delta_1$ and $P_d$. Analysis of the underlying time-domain data yields a switching rate of $1\,$kHz in Fig.~\ref{fig:Region}(d) and $15\,$kHz in (f). In (d) and (e) the histograms show clearly separated states while in (f) the states are connected by faint lines. These lines indicate enhancement of the stochastic switching between the steady states. When the system switching rate becomes comparable to the sampling rate, $f_{sw}\sim f_s$, the states are averaged together.

\begin{figure}
	\center
	\includegraphics{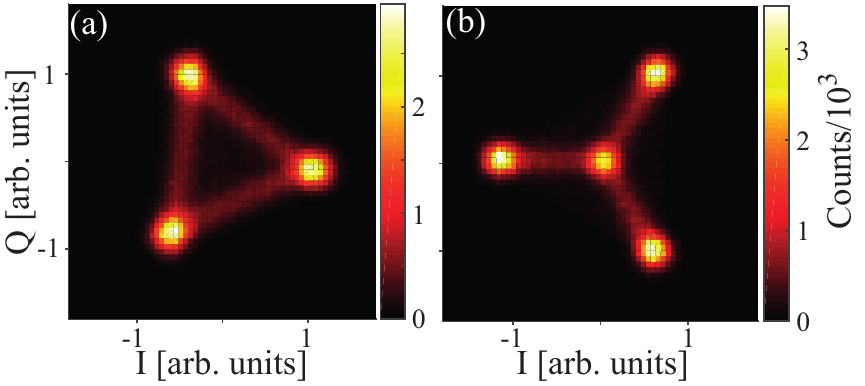}
	\caption{\label{fig:fs10kHz} Histograms sampled with a rate $f_s=10\,$kHz, ten times slower than in Fig.~\ref{fig:Region}(c-f). Here the lines between the steady states are clearly seen. (a) The histogram measured well inside region II reveals a triangle configuration of transitions between the excited state features. (b) The histogram measured at the border between region II and III exhibits a ``star" configuration of transitions between the ground and excited states. These histograms are measured at $\delta_1/2\pi=1.5\,$MHz and drive power as indicated by the white square markers in Fig.~\ref{fig:Region}(a).		
}
\end{figure}

Lowering the sampling rate to $10\,$kHz, makes the switching processes between the stationary states more visible, see Fig.~\ref{fig:fs10kHz}. In region II the transitions occur only between the excited states forming a triangle configuration, while at the border of regions II and III the transitions connect the ground and excited states forming a star configuration.

\subsection{Theory}

To explain the experimental observations and establish a basis for quantitative comparison, we perform a theoretical analysis based on the theory for two-mode resonant dynamics in a frequency-tunable resonator\cite{Wustmann2013,Wustmann2017}. The two-mode equations for slowly varying Heisenberg operators of the coupled modes, $a_1$ and $a_2$, in the doubly rotating frame with frequencies $\omega$ and $3\omega$, have the form
\begin{eqnarray}\label{EOMtext}
 &&i\dot a_1 + (\delta_1 + i\Gamma_1  + \alpha_1 a_1^\dag a_1 + 2\alpha \,a_2^\dag a_2) a_1 + \tilde\alpha \, 
 a_1^{\dag 2} a_2 =  0 
\nonumber\\
&&  i\dot a_2 + (\delta_2 + i\Gamma_2 + \alpha_2\, a_2^\dag a_2 + 2\alpha\, a_1^\dag a_1 ) a_2 + {\tilde\alpha\over 3} 
\,a_1^{3}  \nonumber\\
&&\hspace{5cm}  = \sqrt{2\Gamma_{2,ext}} \, B_2 \,.
\end{eqnarray}
Here, the amplitude of the fundamental mode, $a_1$, describes the subharmonic oscillator, while that of the second mode, $a_2$, acts as an effective parametric pump. $B_2$ is the complex amplitude of the external drive:  
$\Gamma_n$ is the mode damping. Explicit equations for the external damping, $\Gamma_{2,ext}$, and the Kerr coefficients, $\alpha_n$, are presented in \Eqs{Gammas}-(\ref{alphas}). The cross-Kerr coefficients are related to the Kerr coefficients, 
$\alpha=\sqrt{\alpha_1\alpha_2}$ and $\tilde\alpha=\sqrt[4]{\alpha_1^3\alpha_2}$. 
Eqs. \eqref{EOMtext} are associated with and can be derive from the quantum Hamiltonian,
\begin{eqnarray}\label{QHamiltonian}
&& {H}/\hbar =  -\sum_{n=1,2} \left(\delta_n {a}_n^\dag {a}_n + {\alpha_n\over 2}{a}_n^{\dag 2} {a}_n^2 \right) - 2\alpha \,{a}_1^\dag {a}_1 {a}_2^\dag {a}_2  \nonumber\\
 && - {\tilde\alpha\over 3}\left({a}_1^{\dag 3}{a}_2 + {a}_1^{ 3}{a}_2^\dag \right) + \sqrt{2\Gamma_{2,ext}}\left(B_2{a}_2^{\dag} + B_2^* {a}_2\right). 
\end{eqnarray}
Subharmonic oscillations are essentially a classical phenomenon since a large number of photons is generated in the resonator. Thus we restrict the analysis to the quasiclassical solutions of \Eq{EOMtext}, neglecting quantum effects. A detailed description of the 
eigenfunctions and tunneling rates can be found in Ref.~\cite{Zhang2017} for the Hamiltonian \eqref{QHamiltonian} in the single-mode case.

 		
The trivial quasiclassical solution to \Eq{EOMtext}, $a_1=0$, describes a silent oscillator state. It is always stable, see Appendix, \Eq{stability1}. The nontrivial solutions describing stable steady states of the excited oscillator consist of a phase-degenerate triad, the states being stable within the region of existence, \Eq{lambda}. In terms of a polar parametrization of the quasiclassical field amplitudes,
\begin{eqnarray}\label{parametrization}
a_1 = r_1e^{i\phi_1}, \quad  a_2 = {r_2\over \beta}  e^{i\phi_2}, \quad \theta = 3\phi_1- \phi_2  \,,
\end{eqnarray}
with $\beta = \sqrt[4]{\alpha_2/ \alpha_1}$, 
the stable solution has the form,
\begin{eqnarray}\label{r1text}
r_1^2 = {|\delta_1|\over\alpha_1} - {3r_2^2\over 2} + \sqrt{ {r_2^2|\delta_1|\over\alpha_1}  - {7 r_2^4\over4} -  
{\Gamma_1^2\over \alpha_1^2}}\,,
\end{eqnarray}
\begin{eqnarray}\label{sincos_text}
 \sin\theta ={\Gamma_1 \over \alpha_1 r_2r_1}, \quad 
 \theta \in ( \pi/2,\; \pi) \;{\rm mod} (2\pi) \,.
\end{eqnarray}

The solution \eqref{r1text} exists within an interval of the effective pump intensity, $r_{2\pm}^2$,
\begin{eqnarray}\label{rpm_text}
r_2^2 \in (r_{2-}^2 , \; r_{2+}^2), \quad r_{2\pm}^2 = {2|\delta_1| \over 7\alpha_1} \left[1 \pm \sqrt {1 - {7 \Gamma_1^2\over \delta_1^2}} \,\right] ,
\end{eqnarray}
and at negative, red detuning from the fundamental resonator mode,
\begin{eqnarray}\label{threshold_text}
\delta_1 \;\leq \;  - \sqrt7 \,{\Gamma_1 }. 
\end{eqnarray}
The solution \eqref{r1text} is finite at the boundaries of existence \eqref{rpm_text}, \textit{i.e.}, the subharmonic oscillations emerge abruptly when the boundaries  are crossed. The oscillations achieve a maximum intensity,
\begin{eqnarray}\label{r1max_text}
r_{1,max}^2 = {4\over 7\alpha_1} \left(|\delta_1|+ \sqrt{|\delta_1| ^2 - 7\Gamma_1^2}\right)\,, 
\end{eqnarray}
that grows linearly with the detuning far from the threshold, $|\delta_1|\gg \Gamma_1$. The maximum is achieved in this region at  $r_2^2 = |\delta_1|/ 14\alpha_1$. 

The effective pump strength, $r_2$, is defined by a nonlinear response to the external drive $B_2$, \Eq{EOMclass2}. The response exhibits instability at a weak drive, $|B_{2}|^2 \lesssim \beta^2|\delta_1|^3/(18\Gamma_{2,ext}\alpha_1)$, as shown in Figs. \ref{r2_vs_delta} and \ref{r2_vs_B}, but has a regular monostable behavior at larger drive, up to the maximum value given by \Eq{B2max}, 
\begin{flalign}\label{B2max_text}
\!\!|B_{2,max}|^2 \approx {(3\omega_1-\omega_2)^2 |\delta_1|\over 7\beta^2\Gamma_{2,ext}\alpha_1}, \; \delta_1\gg \Gamma_1, \Phi < 0.4\Phi_0\,.
\end{flalign}

The phase $\phi_2$ is defined by the phase of the drive, $\phi_B$, and for the stable branch and large detuning, $\delta_1 \gg \Gamma_{2}$, it is approximately $\pi$-shifted from the latter (see Appendix).
This situation persists within a wide interval of magnetic flux bias, $0<\Phi<0.4\Phi_0$, as long as the anharmonicity of the resonator spectrum exceeds the detuning, $3\omega_1-\omega_2 \gg \delta_1$, see the inset in Fig~\ref{fig:DCtuning}. 


\subsection{Analysis}
\begin{figure*}
	\center
	\includegraphics{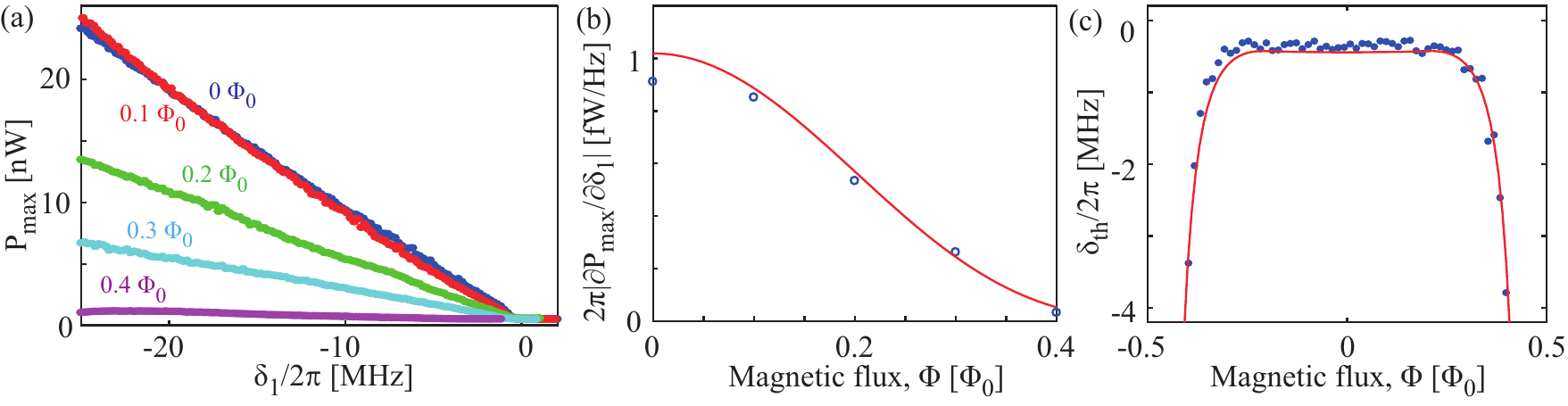}
	\caption{\label{fig:RegionInfo}(a) Growth of the intensity of subharmonic oscillations with red detuning for five different flux bias values. (b) Slopes of the data traces in (a) plotted versus magnetic flux (blue circles);  the red line is the theory prediction, \Eqs{r1max_text} and \eqref{maxima}.  (c) Threshold detunings for subharmonic oscillations at different magnetic flux values; blue dots - experimental data, red line - theory, see comment in the text.
}
\end{figure*}

Using the outlined theoretical results we are able to quantitatively analyze the details in Fig.~\ref{fig:Region}. The (b)-panel displays three linecuts of the subharmonic oscillation region taken at three different values of detuning. The oscillation amplitudes, represented by dots, show sharp onsets at the low-power edge of region II and smoother decays towards the high-power edge. The solid lines correspond to a theoretical fit. The power of the external drive is in linear units
\begin{eqnarray}\label{Pd}
P_d = 3\hbar\omega\, |B_2|^2 10^{Att/10}\,,
\end{eqnarray}
where $Att$ is the attenuation (in dB) between the generator and the resonator for the $3\omega$ drive signal and $|B_2|^2$ is given by Eq.~\eqref{B2r2simple}. The measured output power is
\begin{eqnarray}\label{Pout}
P_{out}=\hbar\omega |a_1|^2 2\Gamma_{1,ext} 10^{G/10}.
\end{eqnarray}
Here $G$ denotes the overall gain of the measurement signal $\omega$ between the resonator and digitizer. The relation between the amplitudes of mode one and two is given in Eq.~\eqref{r1text}. The fit is done by adjusting only one fitting parameter, $X = Q_{2,ext}10^{Att/10}$, which is found in Eqs.~\eqref{Pd} and \eqref{B2r2simple}, where $Q_{2,ext}=\omega_2/2\Gamma_{2,ext}$. The other parameters are measured independently and listed in Table \ref{tab:resonators}; the Kerr coefficient $\alpha_1/2\pi= 85\,$kHz assumes the theory value, and the gain is estimated to be $G=66\pm 0.5\,$dB. The best fit is achieved for $X= (9.97 \pm 0.03) \cdot 10^{11}$. From this we can calculate the photon population of the first resonator mode, $|a_1|^2$. It is found that $2.4\,$nW output power in Fig.~\ref{fig:Region}(a) corresponds to roughly 100 photons.

As seen in Fig.~\ref{fig:Region}(b) the observed oscillations reach a maximum and disappear (the oscillator switches to the ground state) before they reach the theoretical maximum, \Eq{r1max_text}. A comparison of the experimental and theoretical maxima reveals the scaling
\begin{eqnarray}\label{maxima}
|a_{1,max}|^2_{exp} =  0.7 \, |a_{1,max}|^2\,.
\end{eqnarray}
Using \Eq{B2max} and  fitting parameters extracted from Fig.~\ref{fig:Region}(b), we evaluate the boundaries of existence and the stability of the subharmonic oscillations. The upper boundary is presented by the dashed white line in Fig.~\ref{fig:Region}(a). Above the dashed line, in region I, the oscillations do not exist; below this line the theory predicts the existence of oscillations and stability within the whole region II+III (the oscillation lower boundary, \Eq{rpm_text}, lies far below the edge of the panel). However, the oscillations are only visible in the narrow region II but not in III. This can be explained by a competition between the excited states and the stable ground state. At the boundary between regions II and III, the system explores all four available states, as indicated by Fig.~\ref{fig:Region}(d), and in region III the system preferentially stays in the ground state, see Fig.~\ref{fig:Region}(c). Quantitative evaluation of the lower boundary of visibility of the subharmonic oscillations requires a dynamical analysis including the effect of noise, which goes beyond the scope of the present study. 

In Fig.~\ref{fig:Region} all data is taken at zero magnetic flux,  $\Phi = 0$. However, the subharmonic oscillations are detected also at nonzero flux up to $\Phi\approx 0.4\Phi_0$. 
In Fig.~\ref{fig:RegionInfo}(a) we present the maximum output power as a function of detuning for different flux bias values. At $\Phi=0$ this corresponds to the white solid line in Fig.~\ref{fig:Region}(a). The output power is proportional to the maximum population of the first resonator mode and grows linearly with the detuning in good agreement with the theory, \Eq{r1max_text}. Furthermore, the flux dependence of the line slopes in Fig.~\ref{fig:RegionInfo}(a) is in nice agreement with the theory prediction given by the flux dependence of the Kerr coefficient \eqref{alphas}, in \Eq{r1max_text}, and making use of the scaling, \Eq{maxima}, as illustrated in Fig.~\ref{fig:RegionInfo}(b).

The subharmonic oscillations are predicted to start at a threshold at small red detuning, \Eq{threshold_text}. Experimentally, this threshold is defined as the endpoint of the white curve in Fig.~\ref{fig:Region}(a), marked with a white star. Experimental data for the frequency thresholds at different $\Phi$ is presented in Fig.~\ref{fig:RegionInfo}(c) (blue dots). For smaller flux values, $|\Phi| \lesssim 0.4\Phi_0$, the threshold values of the output radiation, \Eq{Pout} evaluated for $\delta_1=-\sqrt{7}\Gamma_1$, exceed the noise level, $P_n \approx 0.44$ nW, and therefore the measurement procedure identifies the true threshold, Eq.~\eqref{threshold_text}. However, at the edges of this region, $|\Phi| \approx 0.4\Phi_0$, the output power rapidly decreases, as indicated in Fig.~\ref{fig:Region}(a)-(b), and therefore the visible oscillation threshold shifts to larger detuning. Quantitatively, the shifted position of the threshold is defined by $P_{out}(|a_{1,max}|^2) = P_n$.  We compute the solution to this equation using the parameters in Table~\ref{tab:resonators}, and the flux dependence of $\Gamma_{1,ext}$ in \Eq{Gammas}. The internal losses and noise power are assumed flux independent. The theory plot of this solution, the red line in Fig.~\ref{fig:RegionInfo}(c), excellently reproduces the data. We note that no fitting parameters were used in Fig.~\ref{fig:RegionInfo}(b) and (c).

\section{Conclusion}

We have observed period-tripling subharmonic oscillations in a driven nonlinear multimode microwave resonator in the quantum regime. When an external drive tone is applied at a frequency $3\omega$, we observe output oscillations at $\omega$, demonstrating period tripling. The output signal consists of three correlated beams having the same amplitudes but with their phases shifted by $2\pi/3$ radians with respect to each other. The oscillations are observed at red detuning from the resonator fundamental mode, and in a finite interval of drive power. Due to the proximity of the second resonator mode to the drive tone, the downconversion efficiency is strongly enhanced, enabling access to the subharmonic oscillation regime. A theory for the two-mode subharmonic resonance was developed to explain the observations. The theoretical predictions are in good quantitative agreement with the experimental observations regarding the boundary of existence of oscillations, maximum output power, and frequency threshold. 
 
Our successful implementation of an intermode interaction of the $a_1^{\dagger 3}a_2$ type may in the future be used to create multi-photon entanglement and multi-component macroscopic cat states \cite{Puri2017}.

\section{Acknowledgments}
We thank Waltraut Wustmann, Mark Dykman and G\"oran Johansson for useful discussions. We gratefully acknowledge financial support from the European Research Council, the European project PROMISCE, the Swedish research council, and the Wallenberg Foundation. J.B. acknowledges partial support by the EU under REA grant agreement no. CIG-618353.

\appendix
\section{}\label{appendix}

In this Appendix we derive quasiclassical solutions to \Eq{EOMtext}, identify the stable solutions, and discuss the solution properties relevant for quantitative interpretation of the experimental data. 

Before proceeding with solving \Eq{EOMtext}, we reproduce the spectral equation for the tunable resonator \cite{Wallquist2006,Wustmann2013} that is used for fitting the data in Fig.~\ref{fig:DCtuning} and justifies the two-mode model for the resonator,
\begin{eqnarray}\label{spectrum}
(k_n d)\tan k_n d  =   \frac{2 E_J (\Phi) }{E_{L,cav}}  - \frac{2 C_J }{C_{cav}}(k_n d)^2\,.
\end{eqnarray}
Here $k_n = \omega_n/v$ is the mode wave vector, $d$ is the length of the resonator, $E_{L,cav}$ is the inductive energy of the resonator; $C_{cav}$ is the resonator capacitance, $C_J\ll C_{cav}$ is the Josephson junction capacitance, and $2E_J (\Phi_{}) = 2E_J \cos(\pi\Phi_{}/\Phi_0)$ is the Josephson energy of the SQUID.  

\begin{figure}
	\centering
	\includegraphics{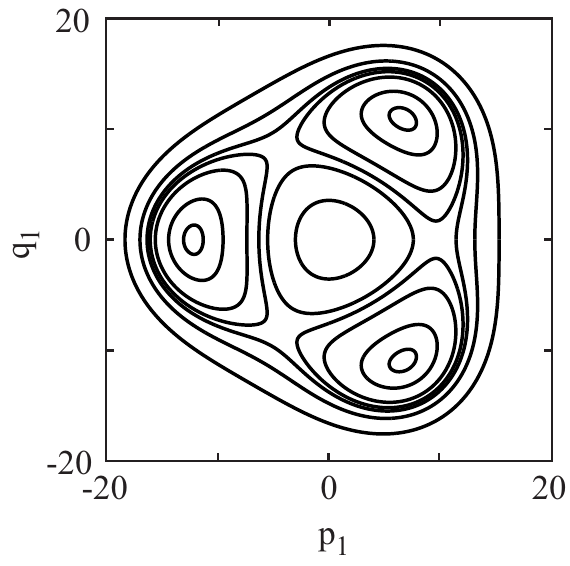}
	\caption{\label{fig:PhasePortrait}Phase portrait for the subharmonic dynamics of the resonator fundamental mode defined by the metapotential, \Eq{Hclass}, with fixed values $p_2$ and $q_2$ given by the experimental data point $\delta_1/2\pi=-12\,$MHz and $|B_2|^2=6.25\cdot 10^{10}\,$photons/s.}
	\label{phaseportrait}	
\end{figure}

It is useful to note that the quasiclassical version of the Hamiltonian \eqref{QHamiltonian}, a metapotential, can be written in  terms of quadratures, $[p_n={\rm Re}(a_n), \,q_n={\rm Im}(a_n)]$, on the form,
\begin{eqnarray}\label{Hclass}
&& H(p_n,q_n)/\hbar  = - \sum_n \left(\delta_n (p_n^2 + q_n^2) + \alpha_n (p_n^2 + q_n^2)^2 \right) \nonumber\\
&& -   2\alpha \,  (p_1^2 + q_1^2)(p_2^2 + q_2^2) \nonumber\\
&&+ {\tilde\alpha\over 3}\left(q_1q_2(q_1^2 - 3p_1^2) - p_1p_2 (p_1^2 - 3q_1^2)\right) .  
\end{eqnarray}
The phase portrait for the period tripling subharmonic oscillator defined by this metapotential is presented in Fig.~\ref{fig:PhasePortrait}.
It gives general information about the structure of the subharmonic oscillator stable steady states: they consist of the four states including the trivial ground state at the origin, $p_1=q_1=0$, and the three nontrivial states corresponding to the excited oscillator.

To establish stability of the trivial solution to \Eq{EOMtext}, $a_1=0$, we linearizing this equation and assume time dependence of small fluctuation, $a_1\propto e^{\lambda_0 t}$, then we find,
 \begin{eqnarray}\label{stability1}
 \lambda_0 = i(\delta_1+ 2\alpha |a_2|^2) - \Gamma_1. 
\end{eqnarray}
Since ${\rm Re} \,\lambda_0 < 0$ the trivial solution is always stable.

Solving \Eq{EOMtext} consists of two steps. First, a solution for the subharmonic oscillations of the first mode is constructed treating the field of the second mode as an effective pump \cite{Hayashi,JordanSmith}. Then the field of the second mode is computed as a nonlinear response to the drive.  Analysis of \Eq{EOMtext} is convenient to perform using dimensionless parameters,
\begin{flalign}
 \!\!\! \delta = {\delta_1\over \alpha_1}, \; 
\Delta = {3\omega_1 - \omega_2\over \alpha_1}, \; 
{\delta_2\over \alpha_1} = 3\delta + \Delta, \; 
 \gamma_n = {\Gamma_n\over \alpha_1}.  
\end{flalign}
Derivation of the explicit equations for the external damping,
\begin{eqnarray}\label{Gammas}
\Gamma_{n,ext} &=& \omega_n (k_nd)\left({C_c\over C_{cav}}\right)^2,  
\end{eqnarray}
where $C_c$ is the coupling capacitance, and for the Kerr coefficients,
\begin{eqnarray}\label{alphas}
\alpha_n &=&  {\hbar \omega_n^2 E_{L,cav}^2\over 16 E_J^3(\Phi) } \,. 
\end{eqnarray}
are found in \cite{Wustmann2013}.  With these parameters, and using the representation (\ref{parametrization}), the stationary \Eq{EOMtext} takes the form, 
\begin{eqnarray}\label{EOMclass}
&& (\delta + i\gamma_1 + r_1^2 + 2r_2^2)r_1 + r_2 r_1^2 e^{-i(3\phi_1 - \phi_2)} =  0 \\
&&  [3\delta + \Delta + i\gamma_2 + \beta^2 (r_2^2 + 2r_1^2)] r_2  + {\beta^2\over 3}r_1^{3} e^{i(3\phi_1 - \phi_2)} \nonumber\\
&& = \beta\sqrt{2\gamma_{2,ext}\over\alpha_1} \,B_2 e^{-i\phi_2} \,. \label{EOMclass2}
\end{eqnarray}
To solve Eq.~\eqref{EOMclass}, we separate the real and imaginary parts,
\begin{eqnarray}\label{ReIm}
\gamma_1 &=& r_2r_1 \sin(3\phi_1 - \phi_2)  \nonumber\\
\delta + r_1^2 + 2r_2^2 &=& -  r_2r_1 \cos(3\phi_1 - \phi_2)  \,,
\end{eqnarray}
and eliminate the oscillator phase. Then we get a closed equation for $r_1$, which has solutions,
\begin{flalign}\label{r1}
r_1^2 = - \left(\delta + (3/2)r_2^2\right) \pm \sqrt{ -r_2^2\delta  - (7/4)r_2^4 -  \gamma_1^2}\,.
\end{flalign}
These solutions are restricted to the region defined by \Eqs{rpm_text} and (\ref{threshold_text}).
Equations for the phase $\phi_1$, extracted from \Eq{ReIm}, read, 
\begin{eqnarray}\label{sincos}
&& \sin(3\phi_1- \phi_2) ={\gamma_1 \over r_2r_1} > 0,  \\
&& \cos(3\phi_1- \phi_2) = \pm \sqrt{1-  {\gamma_1^2 \over (r_2r_1)^2}} = { -\delta - 2r_2^2 - r_1^2 \over r_2r_1} \,.\nonumber
\end{eqnarray}
The solutions have a three-fold degeneracy: for every given value of the phase, $\phi_2$, 
there are three values of the subharmonic oscillation phase, $\phi_1$,  shifted by $2\pi/3$ radians with respect to each other, see  Fig.~\ref{fig:PhasePortrait}.

\begin{figure*}
	\includegraphics{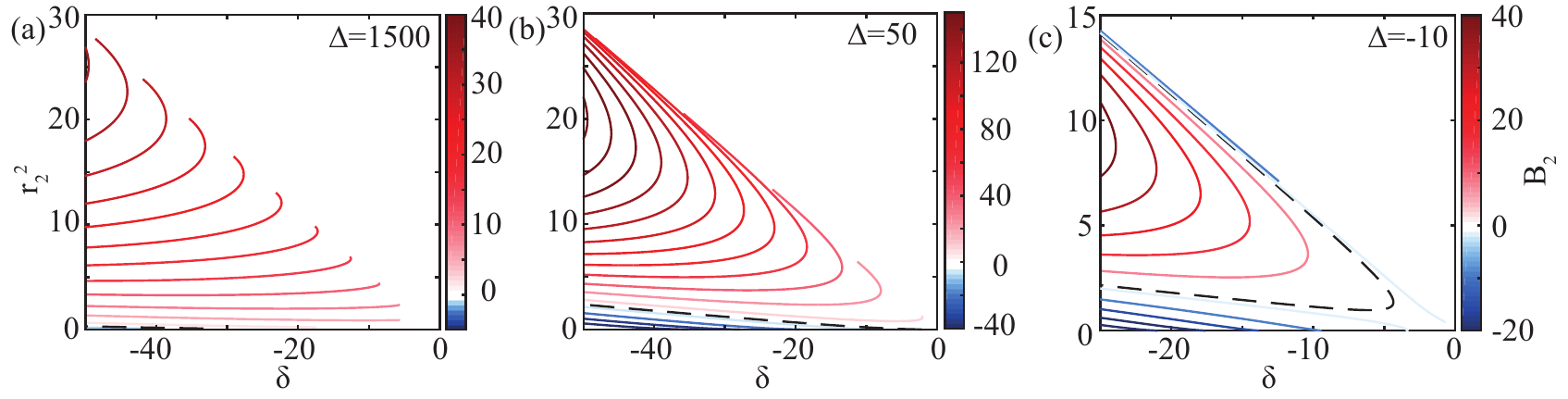}
	\caption{Response of the second mode, $r_2^2$, as a function of detuning for different drive amplitudes $B_2$ (represented by the colour scale). The phase of the response is included in the sign of the drive amplitude. The panels illustrate the evolution of the response with decreasing spectrum anharmonicity. The curves are 
		restricted to the region of existence of subharmonic oscillations. Exact resonance ($B_2=0$) is indicated with a black dashed line. ($\gamma_1 = 1.93, 0.10, 0.08; \; \gamma_2 =0$.)
	}
	\label{r2_vs_delta}
\end{figure*}
To evaluate the stability of these solutions, we use its simplified form, for brevity,  which is valid away from the threshold, 
$|\delta|\gg \gamma_1$,
\begin{eqnarray}\label{ReIm2}
&& 3\phi_1 - \phi_2 = 0, \;\pi  \nonumber\\
&& r_1 = \mp \left({r_2\over 2} \pm \sqrt{|\delta|- {7\over 4}r_2^2}\right) 
\end{eqnarray}
(the -/+ signs in front of the brackets correspond to the $0/\pi$ phase differences). The linearized equation for small fluctuation, $\delta a_1$, around each of the steady state solutions has the form,
\begin{eqnarray}\label{delta_a_1}
i \delta \dot{a}_1 + (\delta + 2r_1^2 + 2 r_2^2) \delta a_1 - (r_1^2 + 2\delta + 4r_2^2) \delta a_1^\ast =0 \,.\nonumber\\
\end{eqnarray}
Assuming, ${\rm Re} (\delta a_1),\, {\rm Im} (\delta a_1) \propto e^{\lambda_1\alpha_1 t}$, for this solution we find from \Eqs{delta_a_1}, (\ref{ReIm2}), 
\begin{eqnarray}\label{lambda}
\lambda_1^2 =  \pm \,6 r_1^2 r_2 ( r_1 \pm \,r_2/2) \,.
\end{eqnarray}
For the lower, minus sign in front of the brackets that corresponds to $3\phi_1-\phi_2 = \pi$ in \Eq{ReIm2}, the exponent is,
\begin{eqnarray}\label{l}
\lambda_1^2 = - \,6 r_2^2 \left( \pm \,\sqrt{|\delta|- {7\over 4}r_2^2} \,\right) \,.
\end{eqnarray}
For the positive root, $\lambda_1^2<0$, hence the solution with both signs positive in \Eq{ReIm2} is stable. The full form of this solution is presented in the main text in \Eqs{r1text}, \eqref{sincos_text}. The other choices of the signs result in positive $\lambda_1>0$, hence corresponding to unstable solutions. 

Equation \eqref{EOMclass2} describes a Duffing oscillator perturbed by back-action of the subharmonic oscillator. The imaginary part of this equation defines the phase of the response, $\phi_2$. Similar to \Eq{sincos}, the difference between this phase and the phase of the drive, $\phi_B$, is defined by the damping, $\gamma_2$, and for the major parameter interval of interest, $\Delta, |\delta| \gg \gamma_2$,  phase $\phi_2$ is either close to the phase of the drive or shifted by $\pi$, $\phi_2 - \phi_B \approx 0, \,\pi$ (cf. \Eq{ReIm2}). The amplitude of the response is found from the equation,  
\begin{eqnarray}\label{B2r2}
&& \left[(3\delta + \Delta  + \beta^2 (r_2^2 + 2r_1^2)) r_2  - {\beta^2\over 3}r_1^{3} \right]^2 + \gamma_2^2 r_2^2 \nonumber\\
&& = \beta^2{2\gamma_{2,ext}\over\alpha_1} \,|B_2|^2 \,. 
\end{eqnarray}

The dependence $r_2^2(\delta)$ for different drive amplitudes $B_2$ and flux values is illustrated in Fig.~\ref{r2_vs_delta}. For better clarity the plots are made neglecting damping of the second mode at the left hand side, and including the phase of the response in the sign of the drive amplitude, then positive $B_2>0$ correspond to $\phi_2=\phi_B$, and negative $B_2<0$ correspond to $\phi_2=\phi_B+\pi$. The response qualitatively resembles the one of the Duffing oscillator; the similarity is most pronounced at small values of the spectrum anharmonicity illustrated in Figs.~\ref{r2_vs_delta}(b) and (c) for $\Delta = 50$ and $\Delta=-10$. Here the bi-stability region is seen at $B_2>0$ as well as the exact resonance, $B_2=0$, which is indicated with a black dashed line in Fig.~\ref{r2_vs_delta}(c). The stable solutions correspond to the lower branch at positive $B_2$, and the branch with negative $B_2$ above the resonance. There is, however, a second resonance that appears at smaller values of $r_2$, the states below this resonance line, at negative $B_2$, are unstable. 

\begin{figure}
	\centering
	\includegraphics{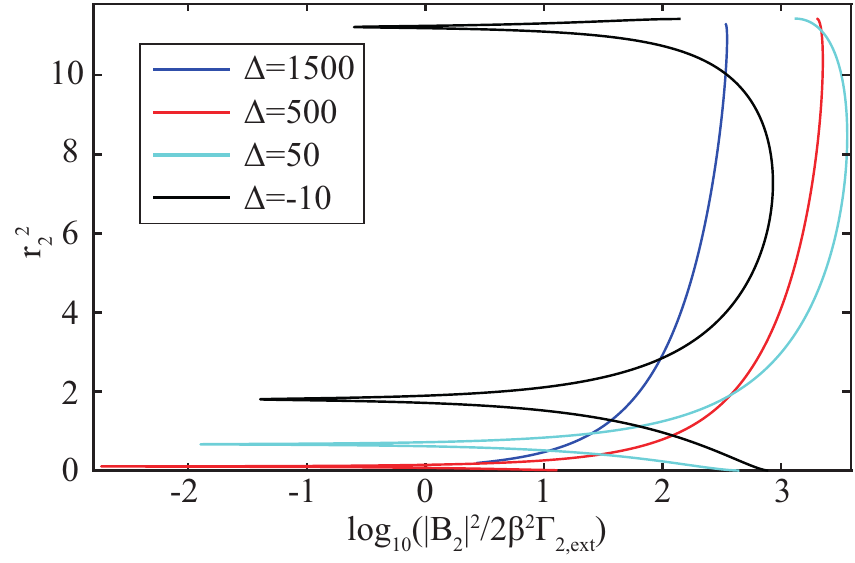}
	\caption{Response of the second mode, $r_2^2$, as a function of drive power for different values of the spectrum anharmonicity,  $\Delta = 1500,  500, 50, -10$, at $\delta =-20$, corresponding to $\Phi=0.1,0.32,0.406,0.4178\,\Phi_0$ ($\gamma_1 = 1.93, 0.43, 0.10, 0.08$, $\gamma_2 = 13.6, 2.94, 0.71, 0.55$, $\gamma_{2,ext} = 13.1, 2.84, 0.69, 0.53$).}
	\label{r2_vs_B}
\end{figure}
The dependence $r_2^2(|B|^2)$ from \Eq{EOMclass2} is illustrated in Fig.~\ref{r2_vs_B} for a representative value of the detuning, $\delta = - 20$, and for different values of the spectrum anharmonicity $\Delta$, which is controlled by the bias magnetic flux $\Phi$. When the spectrum anharmonicity is large, $\Delta = 1500, \,500$ ($\Phi =0.1,0.32 \,\Phi_0$), the stable solution for $r_2$ exists for all drive amplitudes except of very small values, where the second, unstable solution appears (this solution corresponds to the region below the second resonance in Fig.~\ref{r2_vs_delta}). In this region of large anharmonicity, which significantly exceeds the experimental interval of detunings, \Eq{EOMclass2} can be significantly simplified by dropping $|\delta| \ll \Delta, \;  r_1^2, r_2^2 \lesssim |\delta| $, 
\begin{eqnarray}\label{B2r2simple}
&& \Delta^2  r_2^2  = {2\beta^2\gamma_{2,ext}\over\alpha_1} \,|B_2|^2 \,. 
\end{eqnarray}
Inserting \Eq{r1max_text} into this equation we obtain the maximum drive power at which the subharmonic oscillations may persist,
\begin{eqnarray}\label{B2max}
|B_2|^2 \approx {2\Delta^2|\delta| \alpha_1 \over 7\beta^2\gamma_{2,ext}}, \quad |\delta|\gg \gamma_1 \,.  
\end{eqnarray}
When the anharmonicity decreases ($\Delta \lesssim 50$, $\Phi \approx 0.41$), an unstable (back-bending) branch emerges at large drive. This  feature is associated with the bifurcation in Fig.~\ref{r2_vs_delta}. This effect should lead to a reduction of the visible part of the subharmonic oscillation region in Fig.~\ref{fig:Region}(a). With further decrease of the anharmonicity the subharmonic oscillations should disappear, at $\Phi\gtrsim 0.4\Phi_0$.\\ 
\\


%

\end{document}